\documentclass[aps,noshowpacs,prl,superscriptaddress,nofootinbib,notitlepage,twocolumn]{revtex4-1}
\usepackage{graphicx}
\usepackage{amssymb,amsfonts,amsmath}
\usepackage{epstopdf} 
\usepackage{color}
\usepackage{url}
\usepackage[T1]{fontenc}
\usepackage{hyperref}

\begin{document}

\title{Interdisciplinarity: A Nobel Opportunity}

\author{Michael Szell}
\affiliation{Department of Network and Data Science, Central European University, 1051 Budapest, Hungary.}
\affiliation{Center for Complex Network Research, Northeastern University, Boston, MA 02115, USA.}
\affiliation{Complexity Science Hub, 1080 Vienna, Austria.}
\affiliation{MTA KRTK Agglomeration and Social Networks Lendulet Research Group, Centre for Economic and Regional Studies, Hungarian Academy of Sciences, Budapest, Hungary}
\author{Yifang Ma}
\affiliation{Northwestern Institute on Complex Systems (NICO) and Kellogg School of Management, Northwestern University, Evanston, IL 60208, USA.}
\affiliation{Key Laboratory of Mathematics, Informatics and Behavioral Semantics, Ministry of Education, Beijing 100191, China.}
\author{Roberta Sinatra}
\email{robertasinatra@gmail.com}
\affiliation{Department of Network and Data Science, Central European University, 1051 Budapest, Hungary.}
\affiliation{Math Department, Central European University, 1051 Budapest, Hungary}
\affiliation{Center for Complex Network Research, Northeastern University, Boston, MA 02115, USA.}
\affiliation{Complexity Science Hub, 1080 Vienna, Austria.}
\affiliation{ISI Foundation, Torino, Italy.}


\maketitle

Chemists are scientists stirring a flask, physicists solve complicated equations on a blackboard, and physicians, wearing a white coat with a stethoscope around their neck, run against the clock to save a patient. As common as these enduring stereotypes are, they are just as outdated. Today, scientists from different disciplines increasingly work together on complex, previously intractable problems. Consequently, interdisciplinary collaborations now sweep most fields of the natural and life sciences, necessary to tackle the world's most challenging problems \cite{ledford2015hww}. Yet, the scientific enterprise continues to be dominated by old stereotypes: Interdisciplinary science is less likely to receive funding \cite{bromham2016irc} and is discriminated at institutional levels \cite{ledford2015hww}. Ample solutions for funders, institutions and publishers have been suggested \cite{brown2015ihc}, but the most visible form of scientific credit has so far been ignored: How interdisciplinary is our award system? To address this question, we explore interdisciplinarity in arguably the most prestigious award in science, the Nobel Prize. 

\begin{figure*}[t]
\centering
\includegraphics[width=0.92\textwidth]{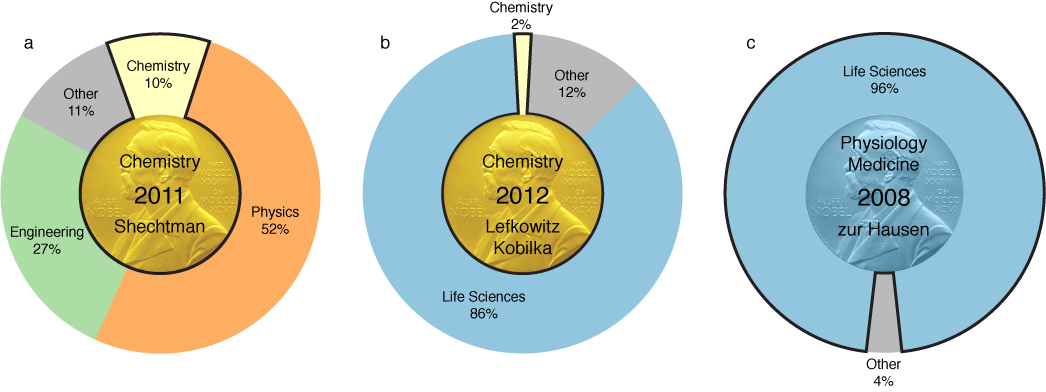}
\caption{{\bf The disciplinary/interdisciplinary impact of Nobel Prize winning discoveries.} a) Shechtman's 1984 paper on quasicrystals \cite{shechtman1984mpl}, rewarded with the Nobel Prize in chemistry in 2011, had a largely interdisciplinary impact, being cited significantly by papers from physics, engineering and its field of award, chemistry. b) Dixon et al.'s 1986 paper on cell receptors \cite{dixon1986cgc}, earning Lefkowitz and Kobilka the prize, is a chemistry Nobel Prize whose impact is almost exclusively outside of chemistry. The paper has been mostly cited by the life sciences. c) In contrast to chemistry, the impact of Nobel Prize papers in physiology/medicine is highly limited to one field -- citations almost exclusively come from the life sciences. A typical example is Schwarz et al.'s 1985 paper on the papillomavirus \cite{schwarz1984sth}, Nobel Prize in physiology/medicine in 2008 to zur Hausen. \label{fig:examples}}
\end{figure*}

In the early 1980s, Dan Shechtman discovered the quasicrystal \cite{shechtman1984mpl}, a regular, but not periodic solid. The discovery that matter could organize itself in \emph{disallowed symmetries} caused an enormous excitement \cite{guardian}, and Shechtman eventually received the Nobel Prize 27 years later. Yet, the award was in chemistry, despite the fact that the discovery of the quasicrystal was published in a physics journal, Physical Review Letters, and had its largest long-term impact in physics \cite{wang2013quantifying}. Indeed, Shechtman's Nobel Prize study has been cited over 3,000 times (Supplementary Information), with 52\% of the citing corpus associated to physics, 27\% to engineering, and only 10\% to chemistry (Fig.\ 1a). Does this mean that Shechtman's discovery was under-appreciated by chemists? The answer is no: Normalizing for the total size of the chemistry literature after 1984, the citations from chemistry were in fact slightly higher than expected by chance. However, the normalized impact on physics and on engineering was around 6 and 2 times higher. Shechtman's paper is therefore a prime example for an interdisciplinary discovery that had a big impact in several disciplines. 

Given that crystallography is on the border of physics and chemistry, the work's interdisciplinary impact is not surprising. However, it makes us wonder: Is Shechtman's award an anomaly, deviating from the expectation that a Nobel Prize should be awarded in the discipline that produced it? To answer this question, we analyzed the interdisciplinary impact of 108 Nobel Prize winning papers \cite{shen2014cca} by all the 59,305 papers that cited them, as recorded by Thomson Reuters Web of Science (Supplementary Information). These Nobel Prize winning papers consist of 25 papers in physiology/medicine (2006-2017), 43 in chemistry (1998-2017), 40 in physics (1995-2017), covering all papers since the Nobel committee started offering a detailed explanation with references for the prize \cite{shen2014cca}. Note that the choices of these years is not ours but comes from the limitation of the data source; a more far-reaching set of Nobel Prize papers or comparable time periods would have been preferred but was not available. 

We find that 60 Nobel Prize discoveries generated very little interest outside of their awarded field. Consider, for example, Schwarz et.\ al.'s 1985 paper on the role of the human papillomavirus in cancer \cite{schwarz1984sth}, acknowledged with a Nobel Prize in physiology/medicine in 2008. The paper received only 41 of its 1,134 total citations from outside of the life sciences (Fig.\ 1c). We do find, however, 35 interdisciplinary discoveries, i.e.~papers that had a big impact in both the awarded and in at least another field. The remaining 13 Nobel Prize winning papers, all awarded in chemistry, are special -- they had only limited impact in chemistry. The prime example is Dixon et al.'s~1986 paper on cell receptors \cite{dixon1986cgc}, the winner of the chemistry prize in 2012, which received 832 of its 984 citations from the life sciences; only 17 came from chemistry-focused journals (Fig.\ 1b).

\begin{figure*}[t]
\centering
\includegraphics[width=0.62\textwidth]{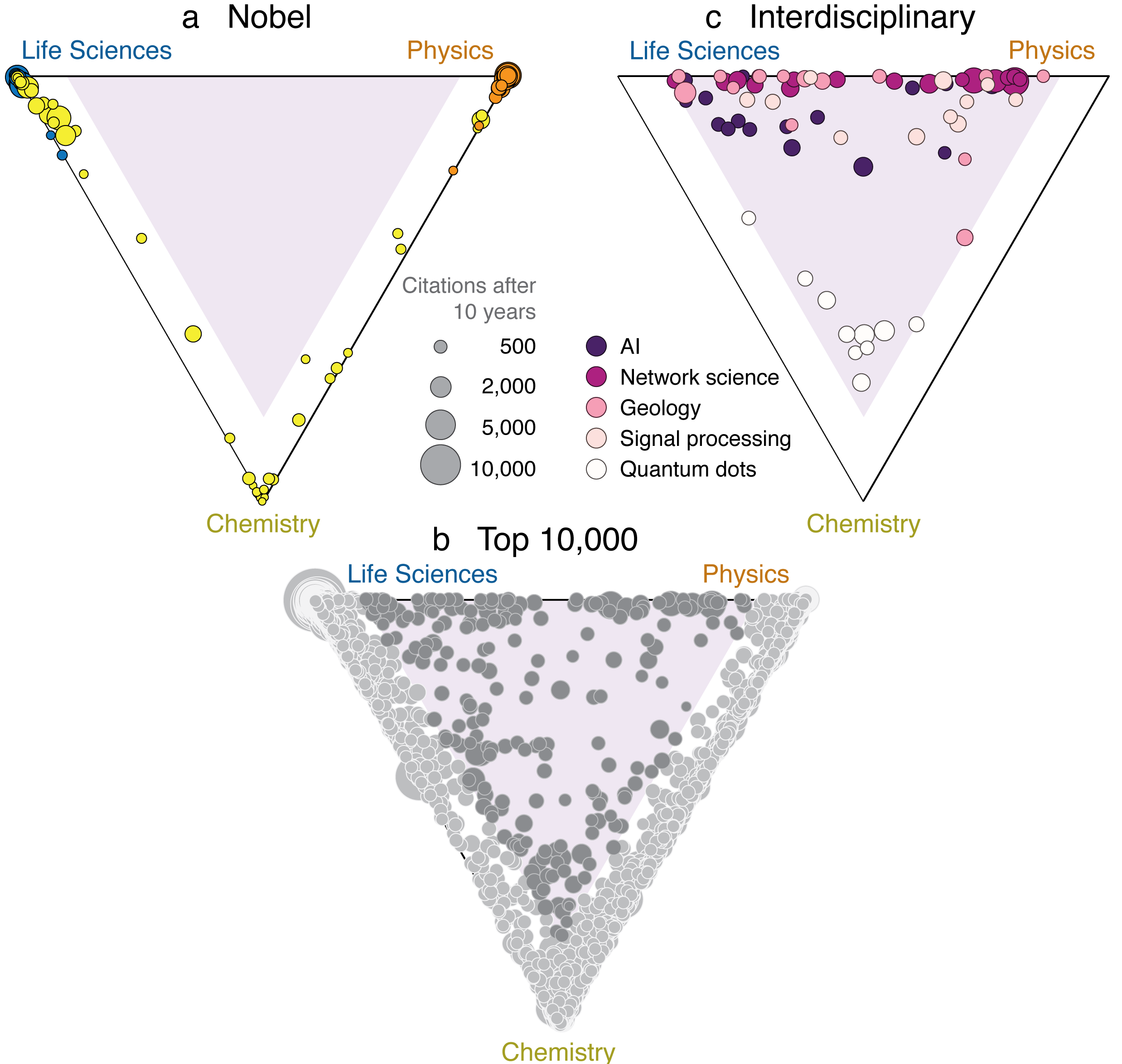}
\caption{{\bf The intellectual space of Nobel Prizes.} Interactive: \url{https://mszell.github.io/nobelplot/nobelplot.html} a) The position of the 108 Nobel Prize winning papers \cite{shen2014cca} in the physics-chemistry-life sciences triangle is determined by how many relative citations each paper received from the respective community. For example, a paper at the center of the triangle received an equal number of citations from all three fields, while a corner position is reserved for papers whose citations came only from one field. Size denotes number of citations after 10 years \cite{wang2013quantifying}, and color denotes field of award: orange physics, yellow chemistry, blue physiology/medicine. The Nobel Prize winning papers are all in a narrow band on the physics-chemistry and the chemistry-life sciences border. No Nobel Prize is awarded to papers in the shaded, interdisciplinary area, especially on the physics-life sciences axis. b) Among the top 10,000 papers in terms of citations after 10 years, only 220 show high degree of interdisciplinary impact, falling into the shaded, interdisciplinary area. c) Of the 220 interdisciplinary impact papers in the shaded area we identify the largest groups by subject: artificial intelligence (AI, 16 papers), network science (16 papers), geology (15 papers), signal processing (11 papers), quantum dots (10 papers).\label{fig:phasespace}}
\end{figure*}

Today, the chemistry prize plays a bridging role in the natural sciences, acknowledging discoveries that either make an impact in chemistry only, that impact both physics and chemistry, or make an impact mostly in the life sciences. Interestingly, most of these cross-disciplinary papers were published after 1980, reflecting the transformation in the field's major research goals from traditional analytical chemistry towards biochemistry \cite{charlton2007tsm,rzhetsky2015choosing} and the emergence of interdisciplinary teams \cite{wuchty2007idt,chemistryworld}. But what about physics and life sciences? Although these fields have fundamentally changed too in the last few decades, for example through increasing interdisciplinary efforts in biological physics, the Nobel Prize in these areas remained deeply disciplinary, as we show below. To understand the degree of interdisciplinarity in the Nobel awards, we plot each Nobel Prize winning paper along a triangle (Fig.\ 2a). A publication is placed on the bottom corner if all the citations of the corresponding paper came from chemistry, likewise the top right corner corresponds to exclusive impact in physics, the top left corner to life sciences. Whenever a paper receives citations from several fields, the paper is placed between the corners, its position reflecting the relative mix of citations. For example, a paper would be at the center of the triangle if it received an equal number of citations from all three fields.

\begin{figure}[t]
\centering
\includegraphics[width=0.42\textwidth]{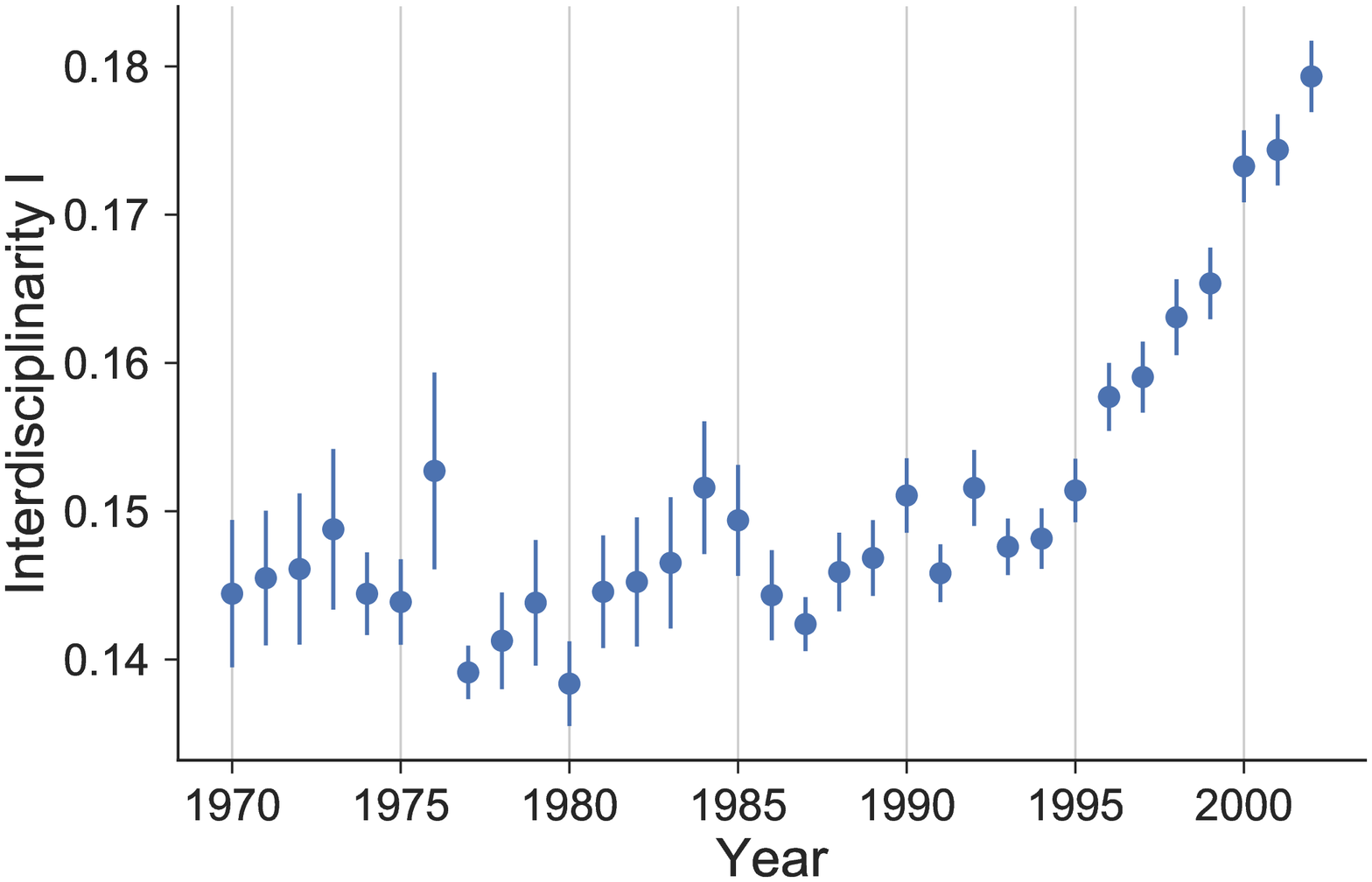}
\caption{{\bf Interdisciplinary research is on the rise.} We define a measure of interdisciplinary impact, $I = 1-G$, using the Gini coefficient $G$, a standard measure for inequality, applied to the number of citations from different fields (Supplementary Information). The value of $I$ ranges from $0$ to $1$. If a paper has $I=1$, then it received an equal amount of citations from each discipline; if $I =0$, it received citations only from one field. The plot shows the evolution of $I$ of the top 10,000 papers from Fig.\ 2b over time, averaged over all papers published each year, errorbars denote s.e.m. Interdisciplinarity of these high impact papers was approximately constant for over two decades but began to rise steadily since the mid 1990s. \label{fig:interdisciplinarityevolution}}
\end{figure}

We make several observations: 1) Papers that receive a Nobel Prize in chemistry (yellow discs) are spread along the chemistry-physics and chemistry-life sciences edges on the triangle, confirming quantitatively the effort by the chemistry prize in acknowledging research that has impact beyond chemistry. 2) In contrast, physiology/medicine Nobel Prize winning papers are all clustered in the narrow vicinity of the life science corner, indicating that they have no impact beyond life sciences. 3) Similarly, most physics prize winning papers are in the narrow vicinity of the physics corner. 4) All Nobel Prize winning papers are located in a narrow band that connects the physics-chemistry and the chemistry-life sciences border. No Nobel Prize is awarded in the shaded area, representing ideas outside this narrow band. In other words, there is no award for work that impacts all three disciplines. 5) In particular, there is a lack of prizes at the physics-life sciences border. 

Could it be that there simply are no high-impact discoveries relevant for both physics and life sciences or to all three disciplines? To answer this question, we plotted the top 10,000 papers in Web of Science in terms of citations after 10 years (Fig.\ 2b). While the Nobel Prize is not given merely for citations \cite{lehmann2006measures,mazloumian2011cbp,radicchi2012reverse,van2014top}, the distribution of top 10,000 papers captures the diversity of ideas important across all fields of science. Indeed, the majority of Nobel Prize winning papers can be found in this top 10,000 list. The plot does confirm the exceptional number of high impact papers at the physics-chemistry and the life sciences-chemistry border, areas regularly awarded by the chemistry prizes. It also shows, however, 220 out of 10,000 papers located inside the interdisciplinary shaded area, documenting the existence of high-impact interdisciplinary discoveries \cite{chen2015tpi} of direct relevance to physics, chemistry, \emph{and} the life sciences. Some of these high impact papers fall onto the physics-life sciences axis, reflecting mostly recent, highly active interdisciplinary areas (Fig.\ 2c) in artificial intelligence (16 papers), network science (16 papers), geology (15 papers), and signal processing (11 papers). Further, we found a cluster of 10 interdisciplinary papers on quantum dots. These fields capture some of the highest impact interdisciplinary areas not yet embraced by the Nobel Prize. 

Taken together, Fig.\ 2\ gives a snapshot of science that is disappointing on two levels. First, despite the understanding that interdisciplinary research is unavoidable in addressing the most challenging problems in current science and society, the vast majority of research is still highly disciplinary. Second, our most prestigious system of scientific recognitions, the Nobel Prize, is reflecting---and possibly cementing---this reality. Although a still relatively small body of interdisciplinary works not being awarded by the prize is statistically not surprising, the fact that only the chemistry prize reaches out towards interdisciplinary subjects flies in the face of the fact that an increasing fraction of high impact discoveries these days have interdisciplinary impact \cite{chen2015tpi,van2015irn}, increasingly between physics and the life sciences. In fact, measuring the interdisciplinarity of papers from the top 10,000 list over time unveils a silver lining. Since the mid 1990s a steady increase of research makes a more balanced impact in different fields (Fig.\ 3), and the time for the Nobel Prize has just arrived to catch up with this reality: Given that it is now 22 years after 1995, when interdisciplinary high-impact papers started to take off, and that the average delay between publication and a Nobel Prize award today is around 20 years \cite{fortunato2014pgt}, we have reached the critical point in time where the issue of recognizing outstanding interdisciplinary research has become pressing (Supplementary Information). 

The Nobel Prize, as most prestigious scientific awards like the Wolf Prize and Lasker Award, was founded to acknowledge advances in a specific discipline. Given that \emph{by definition} these prizes are disciplinary, shouldn't we just leave them alone? The answer is no. First, the Nobel Prize is special, having come to represent to the world science at its best. The few Nobel Prize categories made sense when the prize was established in 1895 \cite{von1981nfr}; science, however, has fundamentally changed since then. Second, the rules of the Nobel Prize have already been generously bent, to award multiple scientists and not just one, and not just for a discovery ``during the preceding year'' as originally stated \cite{yong2017anp}. Since this is the case, why not adapt the prize even further?

The possible unintended consequence of prestigious award systems, like the venerable Nobel Prize, to amplify structural biases, prompts us to ask: Why not create an up to date award system that simply recognizes the best research, rather than pigeonholing findings into specific disciplines \cite{guardian_shakeup}? After all, high impact science is increasingly achieved through the combination of ideas coming from different disciplines \cite{nissani1997tci,rzhetsky2015choosing,van2015irn}. In many ways, interdisciplinarity is happening despite the current reward and support structures that artificially maintain the disciplinary borders within the current scientific enterprise. A renewed system, recognizing research that defies traditional boundaries, could therefore seriously spur innovation towards relatively uncharted territories, as found in the shaded triangle of the studied impact space. To be clear, our point is not that previous Nobel Prizes, or other discipline-focused prizes, were in any way undeserved or should have been given to other discoveries. Our point is more general: Whatever the specific selection processes behind our most prestigious awards may be, they have become out of sync with reality -- and might be suppressing long needed developments \cite{perc2014matthew}. Moving science into the 21st century will only be possible through a rethinking of the traditional boundaries between disciplines \cite{evans2011metaknowledge,sinatra2015cp}, and by making sure that our scientific recognition and credit system is timely \cite{fortunato2014pgt}, open-minded \cite{shen2014cca}, and quantitatively justified \cite{clauset2017dps}.\\

\begin{acknowledgments}
\small{
The authors thank Huawei Shen for providing the data set of Nobel Prize winning papers 9, Federico Musciotto for Web of Science data extraction, and Martina Iori, Federico Battiston, and Albert-László Barabási for helpful comments. RS and MS acknowledge support from AFOSR grant FA9550-15-1-0077 and from the Templeton Foundation grant 61066. RS acknowledges support from AFOSR grant FA9550-15-1-0364 and from the Central European University Intellectual Themes Initiative ``Just Data''.}
\end{acknowledgments}

\bibliography{nobelbib}
\bibliographystyle{Science}

\end{document}


\title{Interdisciplinarity: A Nobel Opportunity \\ Supporting Information}

\author{
Michael Szell$^{1,2,3,4}$, Yifang Ma$^{5, 6}$, and Roberta Sinatra$^{1,7,8,2,3,*}$
}

\blfootnote{\scriptsize $^1$Department of Network and Data Science, Central European University, 1051 Budapest, Hungary. $^2$Center for Complex Network Research, Northeastern University, Boston, MA 02115, USA. $^3$ Complexity Science Hub, 1080 Vienna, Austria. $^4$MTA KRTK Agglomeration and Social Networks Lendulet Research Group, Centre for Economic and Regional Studies, Hungarian Academy of Sciences, Budapest, Hungary, $^5$Northwestern Institute on Complex Systems (NICO) and Kellogg School of Management, Northwestern University, Evanston, IL 60208, USA. $^6$Key Laboratory of Mathematics, Informatics and Behavioral Semantics, Ministry of Education, Beijing 100191, China. $^7$Math Department, Central European University, 1051 Budapest, Hungary. $^8$ISI Foundation, Torino, Italy. *To whom correspondence should be addressed. Email: robertasinatra@gmail.com}

\baselineskip15pt
\linespread{1.05}

\maketitle

\section{Selection of Nobel Prize winning papers}
Our initial data set consists of the Nobel Prize winning papers identified in Ref.~\cite{shen2014cca}, covering all papers up to 2013 since the Nobel committee started offering a detailed explanation with references for the prize, and additionally the Nobel Prize winning papers of the years 2014, 2015, 2016, and 2017 identified with the information provided on the nobelprize.org page. Of all these papers, we were not able to consider the following three types: 1) Papers of 4 laureates that are not contained in the Web of Science database, for example Tu et al.\ (1981), Acta pharmaceutica Sinica, 16(5), 366-370, 2) 6 papers that were published after 2002, since our citation data is available until 2012 and citations 10 years after publication cannot be determined here, for example Novoselov et al.\ (2004), Science 306, 666-669, and 3) 17 papers that were sleeping beauties, i.e.\ papers that had an insufficient number of citations 10 years after publication from life sciences, physics and chemistry to be placed in the triangle of Fig.\ 2.\ with sufficient statistical confidence, for example Kroemer (1996), Physica Scripta T68, 32-45, which had only 5 citations 10 years after publication. These discarded, and the resulting 108 Nobel Prize winning papers, their identifiers and bibliometric data, are provided in the SI Data file. Citations from different fields can be fractional numbers when citing papers belong to multiple disciplines. Paper disciplines are identified by using the journal subject categories of Web of Science. Note that we did not consider Nobel Prizes in economics because of the sparsity of economics papers in the Web of Science data set. The dataset is available at \cite{dataset1}.

\section{Selection of top 10,000 papers}
The top 10,000 papers shown in Fig.\ 2b are the 10,000 papers in the whole Web of Science data set that had the most citations 10 years after publication, published before 2002 for the same reason as above.

\section{Validation of results with a second data set}
\begin{figure*}[t]
\centering
\includegraphics[width=0.28\textwidth]{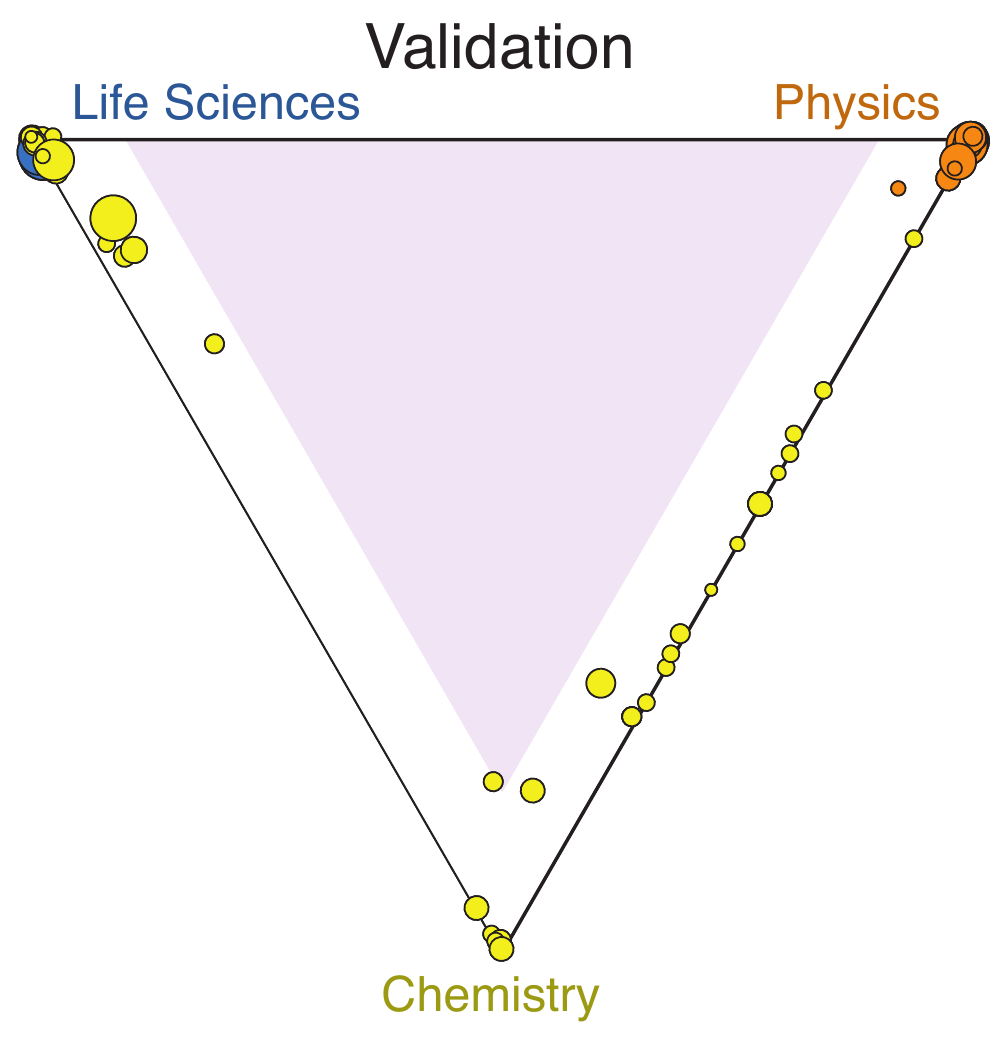}
\caption{{\bf Validation of results with a second data set.} Using the 99 papers of Nobel laureates that were published in their year of ``key research'' (data  from Ref.\ \cite{jones2011age}) shows the same results as Fig.~2.\label{fig:validation}}
\end{figure*}

Because Nobel Prize winning papers are sometimes inconsistently or not explicitly identified on the nobelprize.org website, we cross-checked our results with another data set in which the key works of Nobel Prize winners were manually curated \cite{jones2011age}, available from http://www.nber.org/nobel/. Using the year of ``key research'' from this data set \cite{jones2011age}, we identified in the Web of Science database for each Nobel laureate (who is part of our data set) all papers in that year. Due to name disambiguation issues we manually removed the papers that were wrongly attributed in the Web of Science search query to a Nobel laureate; for example papers of Diana M.\ Lee (attributed to laureate David M.\ Lee) or Roger J.\ Davis (attributed to laureate Raymond Davis Jr.). Finally, for each Nobel laureate we selected their most cited paper in that year. The resulting 99 papers are visualized in Fig.\ \ref{fig:validation}. This second data set is not perfectly overlapping with our main data set, nevertheless the conclusions are identical.

\section{Increasing number of papers and interdisciplinarity}
A substantial time delay exists between the date of publication of a Nobel Prize paper and the date of its Nobel Prize award. This delay has increased over time, having typically reached on average around 20 years \cite{fortunato2014pgt}. Given the increase of interdisciplinary impact over time, Fig.\ 3, we ask: is our analysis not relevant because of the recency of the papers studied? As we show below the answer is no, as we have already reached a few years ago the boundary of typical time delay where the issue has become truly pressing. 

\begin{figure*}[t]
\centering
\includegraphics[width=0.88\textwidth]{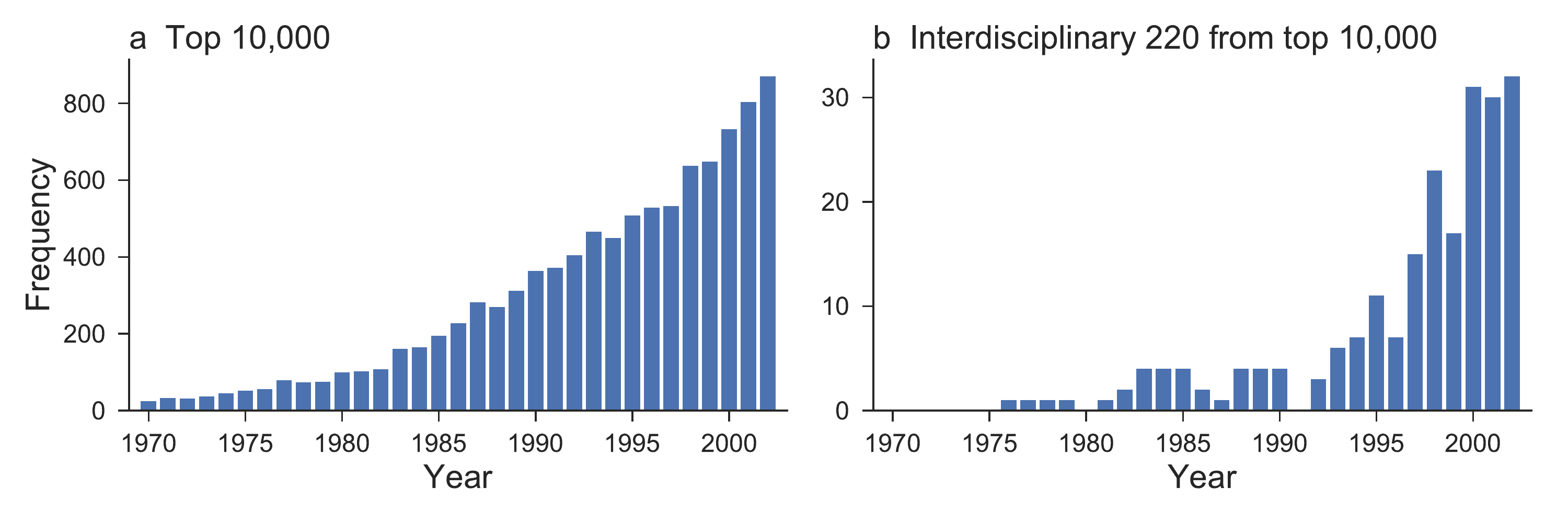}
\caption{{\bf Distribution of publication dates of the top 10,000 papers.} a) The publication years of the 10,000 highest impact papers (in terms of citations after 10 years) follow a right-skewed distribution, with more papers having been published in more recent years, in line with the well documented growth of scientific output over time. b) The distribution for the subset of 220 interdisciplinary papers that fall into the shaded area of Fig.\ 2b, is also right-skewed. Here the skewness is even stronger, with half of the papers coming from the last five years 1998-2002 of the data set, consistent with the increase of interdisciplinarity since the mid 1990s, Fig.\ 3.\label{fig:yearlydistribution}}
\end{figure*}

The 220 interdisciplinary papers that fall into the shaded area of Fig.\ 2 follow a comparable distribution of publication years as all the top 10,000 papers. This comparison is shown in Fig.\ \ref{fig:yearlydistribution}. Consistent with the measured increase of interdisciplinarity since the mid 1990s, Fig.\ 3, the distribution of publication years of the interdisciplinary papers, Fig.\ \ref{fig:yearlydistribution}b, is slightly more right-skewed than the distribution of all 10,000 papers, Fig.\ \ref{fig:yearlydistribution}a. About half of the 220 interdisciplinary papers were published up to 1997, more than 20 years in the past, while the other half was published between 1998 and 2002, now between 15 and 20 years ago. These observations imply that enough time has passed to account for award delays for our dataset of papers to be relevant. The longitudinal analysis demonstrates that, although increased focus on interdisciplinary research is a relatively recent phenomenon, it is now high time to develop an adequate credit system for outstanding scientific accomplishments of any kind.

\section{Measure of interdisciplinarity}
We measure the interdisciplinary impact $I$ of papers as $I = 1-G$, Fig.\ 3, where $G$ is the Gini coefficient, a standard measure for inequality, applied to the number of citations coming from each discipline within 10 years after publication. The value of $I$ ranges from $0$ to $1$. If a paper has $I=1$, then it received an equal amount of citations from each discipline; if $I =0$, it received citations only from one field. All the different disciplines of journals are given by Web of Science as Life Science, Physics, Chemistry, Computers, Engineering, Mathematics, Agriculture \& Environment, Humanity, Others. When the number of fields is restricted to only Life Science, Physics, Chemistry, we find the same result of increased interdisciplinarity since the mid 1990s, only with naturally larger fluctuations, showing robustness of the method. 

\bibliographystyle{Science}
\bibliography{nobelbib}